\documentclass[12pt]{JHEP}
\usepackage{amssymb}
\usepackage{latexsym}
\usepackage{graphicx}
\usepackage{epsfig}
\title{An Effect of $\alpha'$ Corrections on Racetrack Inflation}
\author{Brian R. Greene and Amanda Weltman\\ 
 Institute for Strings, Cosmology and Astroparticle Physics,  \\ 
 Department of Physics, \\ 
 Columbia University, \\ 
 NewYork, \\ 
 NY 10027,\\
  USA \\ 
  \email{greene@phys.columbia.edu} \\ 
  \email{weltman@phys.columbia.edu}} 

\abstract{We study the effects of $ \alpha '$ corrections to the
K\"ahler potential on volume stabilisation and racetrack inflation.
In a region where classical supergravity analysis is justified,
stringy corrections can nevertheless be relevant for correctly
analyzing moduli stabilisation and the onset of inflation.}
\keywords{Inflation, moduli stabilisation}
\begin{document}
\section{Introduction}
Recent studies seeking to embed inflation in string theory in the context of 
Type IIB flux compactifications \cite{GKP, kklt, kklmmt, berg, cline, burgess, RI, per, vj, rev}
have revealed the importance and the delicate nature of moduli
stabilisation:  moduli need to be stabilised with a large enough
mass to avoid Equivalence Principle violations while still allowing
sufficient flat regions in the potential for the onset of inflation.
This suggests the possibility that a successful realization of inflation in
string theory might place interesting constraints on the moduli of the compact manifold.
With the wealth of possible string compactifications, such potential phenomenological
constraints on the parameter space are worthy of study; conversely
with the wealth of proposals in
the literature for instantiating inflation,
any restrictions curtailing the possibilities similarly deserves attention.

In this note, we focus on the specific proposal of
racetrack inflation in string theory \cite{RI}, which has
been studied in the large radius limit of Calabi-Yau compactifications.
An interesting question is whether this approach to inflation can
be realized as we move to smaller radii, or whether corrections
that contribute away from large radius might significantly affect
the analysis. Naively, one might anticipate that so long as we remain
at a radius large enough to justify perturbation theory, such corrections
would have little relevance\footnote{We thank P. Berglund for bringing to our attention 
\cite{per} in which the authors find that stringy corrections are relevant even at large volumes in agreement with the results presented in this note.}
. The requirement of successful inflation, however,
is far more constraining than perturbative reliability and hence even at
moderately large radii, perturbative corrections can be important to
the cosmological analysis.
 
Specificially, we study the effect that the leading
perturbative corrections to the K\"ahler potential have on
cosmological studies using the supergravity scalar F-term potential.
For the cases we study, such corrections can undermine the onset of
inflation in a region of moduli space where classical supergravity
analysis concludes inflation should occur. In some cases, it may be
possible to choose a new set of parameters to yield inflationary
dynamics; even so, the parameters would generally be unrelated to
those studied in the literature, and additionally, the choice would
depend on the specific Calabi-Yau on which one
compactifies as well as the string coupling constant $g_s$. It would be interesting
to include $\alpha'$ corrections at the outset in any search for inflationary potentials as this may provide a tighter handle on the available regions in parameter space in which one could stabilise the moduli and find inflation.  \\

\noindent This note is organised as follows. In section $2$ we quote
the result for the corrected K\"ahler potential and illustrate that
the supersymmetric minimum is insensitive to this correction. By
using the corrected K\"ahler potential, in section $3$, we show that
the no-scale structure is broken and a potential for the volume
modulus is generated. In order to be self-contained, section $4$ is
devoted to a brief review of the relevant results from racetrack
inflation \cite{RI}. The slow roll parameters are computed for the
case of interest. In section $5$ we explore the effect of the
corrections on volume stabilisation and on the conditions necessary
for inflation. In addition, we choose specific compactifications to
study and for each find the minimum value for the volume modulus
that ensures $\alpha'$ corrections can be safely ignored. Our results indicate
that such minima are generally deep in the perturbative domain
and hence $\alpha'$ corrections can be relevant even at reasonably 
large volume.

\section{The Corrected K\"ahler Potential}

The leading perturbative corrections to the K\"ahler potential of
the volume modulus have been computed in \cite{beckers} using the results of 
\cite{afmn, gw, fp, grisaru, can}. With $2\pi
\alpha' = 1$ \cite{pol}, the K\"ahler potential including $\alpha'$
corrections is

\begin{equation}
K= -2\log(\hat{\mathcal{V}} + \frac{1}{2}\xi e^{-3/2 \phi}),
\end{equation}

\noindent where $ \xi = -\frac{1}{2} \chi \zeta (3)$  and $\chi$ is
the Euler number of the compactification manifold. To facilitate
comparisons with earlier results, we work in the Einstein frame;
this is the origin of the dilaton dependance of the correction. While the
result holds for any number of K\"ahler moduli, we will restrict
ourselves to a single K\"ahler modulus, $T$. As such the Calabi-Yau
volume, $\hat{\mathcal{V}}$, is related to the volume modulus, $T$,
by $\hat{\mathcal{V}} = (T + \overline{T})^{3/2}$. In keeping with
\cite{kklt} and \cite{RI}, we treat the complex structure moduli and
the dilaton as having been stabilized prior to the fixing of the
volume modulus. It is clear from the above, however, that one should
include stringy corrections prior to this fixing as all fields
should be stabilized using the corrected K\"ahler potential. At the
SUSY minimum

\begin{equation}
D_iW = \partial_iW + W \partial_i K = 0,
\end{equation}

\noindent where $i$ runs over the complex structure moduli and the
dilaton. This should be contrasted with

\begin{equation}
D^{(0)}_iW = \partial_iW + W \partial_i K^{(0)} = 0,
\end{equation}

\noindent
where $K^{(0)} $ is the tree level K\"ahler potential.
Fixing moduli using $(2)$ rather than $(1)$ is expected to
change the specific values where the complex structure
moduli and the dilaton are fixed but not the systematics.
Hence the correction term can be treated as a constant
for any specific Calabi-Yau. Henceforth,
we set
$L = -\frac{1}{4} \chi \zeta (3)e^{-3/2 \phi} = - \frac{1}{4}\chi \zeta(3) g_s^{-3/2 }$. \\

\noindent At the SUSY minimum, the potential is insensitive to the
$\alpha' $ corrections as one would expect from non-renormalisation
theorem arguments \cite{nonrenorm}. It is instructive to see how this happens explicitly
here. At the SUSY minimum

\begin{equation}
D_TW = 0 \rightarrow W = -\frac{\partial_T W}{\partial_T K} =  -\frac{(\partial_T W)((T+ \overline{T})^{3/2} + L)}{( -3 (T+ \overline{T})^{1/2})}
\end{equation}

\noindent The scalar potential,  $V = e^K ( g^{T\overline{T}} D_TW
\overline{D_TW} - 3|W|^2)$ at the minimum is

\begin{eqnarray}
V_{SUSY}& =& -3 e^K |W|^2  \nonumber  \\
& = & -3 ((T+ \overline{T} )^{3/2} + L)^{-2} ((T+ \overline{T})^{3/2} + L)^2 \Bigl(\frac{(\partial_T W)^2}{( 9 (T+ \overline{T})} \Bigr) \nonumber \\
 &=& -\frac{(\partial_T W)^2}{3 (T+ \overline{T})}
\end{eqnarray}

\noindent We see from this how the  correction term drops out in the
final result. Of course, this will no longer be the case away from
the SUSY minimum.

\section{Breaking the No-scale Structure}

\noindent The classical supergravity potential  displays a no-scale
structure which makes fixing the volume modulus more subtle. In
\cite{kklt} this fixing was achieved by including non-perturbative
corrections to the superpotential that break the no-scale structure
and generate a potential for $T$. One can, however, break the
no-scale structure in other ways. In particular, $\alpha'$
corrections to the K\"ahler potential break this structure and
generate a correction to the supergravity potential dependant on $T$
and proportional to the Euler number of the internal manifold. With
the corrected K\"ahler potential and its metric, $g_{T\overline{T}}
= [3(T+\overline{T}) - 3/2L
(T+\overline{T})^{1/2}]/[((T+\overline{T})^{3/2}+L)^2]$ inserted
into the supergravity potential

\begin{equation}
V_F = e^K \Big(g^{T\overline{T}}D_TW \overline{D_TW} - 3|W|^2 \Big)
\end{equation}

\noindent it can be seen that the two terms no  longer cancel, and a
potential for the volume modulus, $T$, is generated. Since we only
consider tree level contributions to the superpotential arising from
the fluxes, $W=W_0$ here we find

\begin{equation}
 V_T = \frac{3LW_0^2}{(2(T+\overline{T})^{3/2} - L)}.
\end{equation}

\noindent As expected, the $L=0$  limit gives the expected $V_T=0$
result. Alone, however, the stringy corrections to the K\"ahler
potential cannot stabilise the volume modulus as the potential
exhibits runaway behaviour \cite{beckers}.

\section{Review of Racetrack Inflation}

From a cosmological standpoint, simply finding a potential for the
volume modulus is not sufficient. One needs to find a minimum where
$T$ can be stabilised and, importantly, with a potential that
exhibits a sufficiently flat
region along which inflation can occur. \\

\noindent In KKLT \cite{kklt}, the no-scale structure is broken by
adding non-perturbative corrections to the superpotential and the
resulting AdS minimum is subsequently lifted to a dS minimum by
adding a stack of antibranes thus breaking supersymmetry.  In
\cite{RI}, the authors include additional non-perturbative
potentials of the modified racetrack type and find saddle point
regions in the potential where slow-roll inflation can take place.
In this sense the authors of \cite{RI} have revived the old ideas of
modular inflation \cite{bg, banks} with a flat enough potential. In
the case of \cite{RI} the inflaton will be $Y = \Im (T)$.

\noindent The superpotential considered includes  non-perturbative
corrections and is assumed to have the modified racetrack form,

\begin{equation}
W=W_0+A\,e^{-aT}+B\,e^{-bT} ,
\end{equation}

\noindent where $W_0$ is the effective superpotential as a function
of all the complex structure moduli and the dilaton, all assumed to
have already been stabilized. As in the KKLT scenario, $W_0$ is
required to be small ($W_0 \lesssim 10^{-4}$). This is achieved by
discretely tuning fluxes. The exponential terms are expected to
arise from gaugino condensation in a theory with a product gauge
group \cite{gprod}. For example, for an $SU(N) \times SU(M)$ gauge group
one finds $a=2\pi/M$ and $b=2\pi /N$. $A$ and $B$ are expected to be
small in Planck
units \cite{gprod}.\\

\noindent
The scalar potential receives contributions from two terms\footnote{In \cite{west} an inflationary potential is claimed to be found using $\alpha'$ corrections instead of an uplifting potential term.} 

\begin{equation}
V = V_F +\delta V.
\end{equation}

\noindent The first is the standard $\mathcal{N} = 1$ supergravity
F-term potential, $(6)$, \cite{noscale} and the second term, $\delta
V$, is induced by the tension of the anti-D3 branes added to break
supersymmetry and lift the potential from an AdS to a dS minimum
\cite{kklt} . Conveniently, the introduction of the anti-branes does
not introduce extra translational moduli as their position is fixed
by the fluxes. Their contribution is positive definitive and is of
the form \cite{kpv}

\begin{equation}
\delta V = \frac{E}{X^\alpha},
\end{equation}

\noindent where the coefficient $E$ depends on the the  tension of
the branes $T_3$, the number of branes and the warp factor. For this
reason one can discretely tune $E$ and the supersymmetry breaking in
the system but not to arbitrary precision. In \cite{RI} $E$ is tuned
to set the global minimum at the Minkowski vacuum, $V=0$.  In
\cite{kklt} a metastable de Sitter solution is found by tuning $E$
so that the minimum is dS with a small cosmological constant.
Depending on the location of the anti-branes one finds different
results for the exponent $\alpha$. If the anti-branes are situated
at the bottom of the throat in the region of maximum warping one
finds $\alpha = 2$. On the other hand if the anti-branes sit in the
unwarped region $\alpha =3$ \cite{RI}. Since the former is
energetically favoured we set $\alpha=2$ henceforth.

The shape of the resulting potential is  highly sensitive to the
parameter values. Including stringy corrections essentially adds a
new parameter and, we show below, this added term has the capacity
to destabilise some of the key features found in \cite{RI}. The
following set of parameters were chosen by \cite{RI} to illustrate
the presence of a saddle point region where $X= \Re(T)$ has a
minimum and can thus be stabilised, and $Y= \Im(T)$ has a
sufficently flat maximum to ensure that a field starting at the
saddle point and rolling in the $Y$ direction will yield inflation.

\begin{equation}
A = \frac{1}{50}, \quad B= \frac{-35}{1000}, \quad a = \frac{2 \pi}{100}, \quad B= \frac{2 \pi}{90}, \quad W_0 = \frac{-1}{25000}, \quad E = 4.14668 \times 10^{-12}
\end{equation}

\begin{center}
\EPSFIGURE[h]{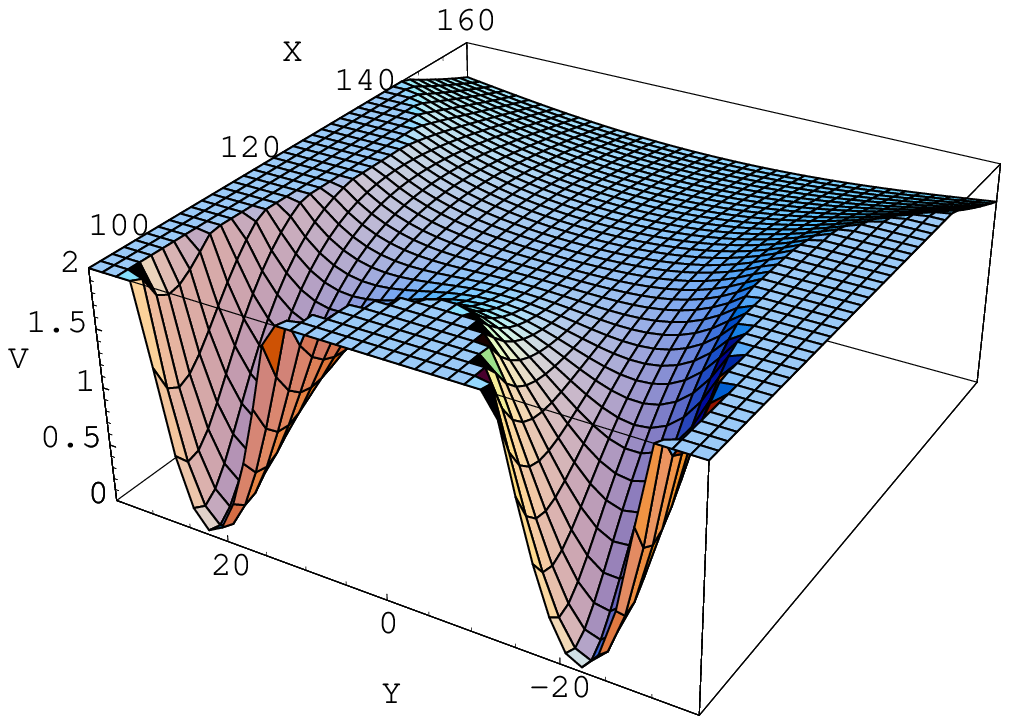}{ Original Racetrack potential corresponding to $L=0$, (rescaled by $10^{16}$)}
\end{center}

The potential has a saddle point at

\begin{equation}
X_{\rm sad} = 123.216, \quad Y_{\rm sad}=0, \quad V_{\rm saddle} = 1.655 \times 10^{-16}
\end{equation}

and minima at

\begin{equation}
X_{\rm min} = 96.130, \quad Y_{\rm min} =\pm 22.146
\end{equation}

In Figure 2, a plot of the $Y=0$ slice of the  potential is included
to illustrate the crucial minimum in the $X$ direction.

\begin{center}
\EPSFIGURE[h]{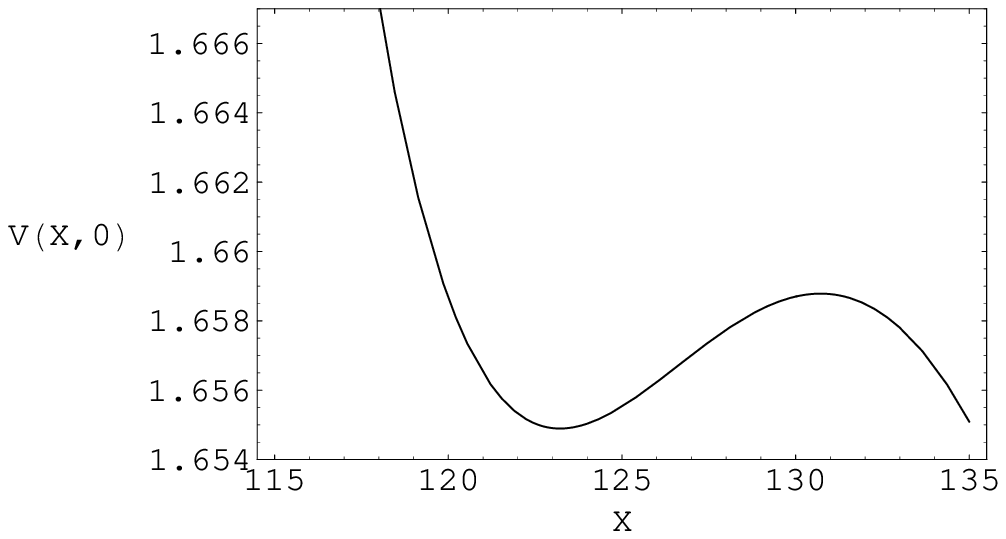}{$Y=0$ slice of original Racetrack potential for $L=0$, (rescaled by $10^{16}$) }
\end{center}

\subsection{Computing Slow Roll Parameters}

We illustrate the claim that one can  get slow roll inflation near
the saddle point for the racetrack potential.  For successful
slow-roll inflation the following conditions need to be met (in
units where $M_P = 1$)\footnote{We thank L. McAllister for pointing out a typo in a previous version.}

\begin{equation}
\epsilon \equiv \frac{1}{2}\Bigr(\frac{V'}{V} \Bigl)^2 \ll 1 \, , \quad \eta \equiv \frac{V''}{V} \ll 1,
\end{equation}

where the primes refer to derivatives  with respect to a canonically
normalised scalar field. Since $V' = 0$ at the saddle point,
$\epsilon$ is exactly zero. To compute $\eta$ one must take account of
the non-canonical kinetic term for the inflaton, $Y$. It is useful
to derive $\eta$ in general so that the result can be used when we
include the correction term. Specifically the kinetic term is

\begin{equation}
{\cal L}_{\rm kin} = 2g_{T\overline{T}} \frac{1}{2} (\partial_\mu X \partial^\mu X +  \partial_\mu Y \partial^\mu Y) ,
\end{equation}

\noindent where the factor of 2 is a  result of the relation between
$X$ and $T$. Taking this normalisation into account yields

\begin{equation}
\eta = \frac{ V''}{V}  \, \rightarrow \,  \eta = \frac{ V''}{2g_{T\overline{T}}V}.
\end{equation}

For the large  radius case, $g_{T\overline{T} } = 3/(4X^2)$,
resulting in $\eta = [2X^2 V'']/[3V]$ with $X$ evaluated at the
saddle point and we find.

\begin{equation}
\eta_{\rm saddle} = -0.0061.
\end{equation}

\noindent This agrees with \cite{RI}.\footnote{We thank J.J. Blanco
Pillado for allowing us to compare results exactly. } It is now a
simple matter to include the effect of the loop term and study its
effect on $\eta$. Specifically

\begin{eqnarray}
\eta &=& \frac{1}{2g_{T \overline{T}}} \frac{V''}{V} \nonumber \\
&=& \frac{V''{((T+\overline{T})^{3/2}+L)^2}}{3V(2(T+\overline{T}) - L (T+\overline{T})^{1/2})} \nonumber \\
&=& \frac{V''{((2X)^{3/2}+L)^2}}{3V(4X - L (2X)^{1/2})},
\end{eqnarray}

\noindent
where, in the above expressions $V$ also has $L$ dependance.

\section{Effect of $\alpha'$ Corrections}

There are three features  required of the potential, all of which
were met in \cite{RI}.
\begin{itemize}
\item The global minimum must be dS or Minkowski.
This step is achieved by including the $\delta V$ term \cite{kklt}.
One can tune the value of the potential at the minimum by discretely
tuning $E$.
\item A de Sitter saddle point with a minimum in the $X$ direction.
\item A sufficiently flat maximum in the $Y$ direction to
attain a cosmologically significant duration of slow roll inflation.
\end{itemize}

\noindent We find, for any reasonable value of $L$, the stringy
correction spoils the second and third conditions. By adjusting the
parameters it may then be possible to re-establish a minimum, but it
is difficult to do so while keeping $\eta$ sufficiently small.
Considering the amount of fine-tuning required to find slow roll
inflation near the saddle point, this is not unexpected. The
correction we are considering amounts to a new, non-tunable term not
previously considered in this context.
Since some of the parameters in the model
can only be fine tuned discretely, it may be
that for any specific example, one cannot find regions
conducive to inflation.\\

\noindent This possibility can be illustrated by plotting the cross
section through the saddle point for non-zero $L$. Here we choose
$L=60$ which corresponds to the quintic, $\chi = -200$, and string
coupling $g_s \approx 1$.

\begin{center}
\EPSFIGURE[h]{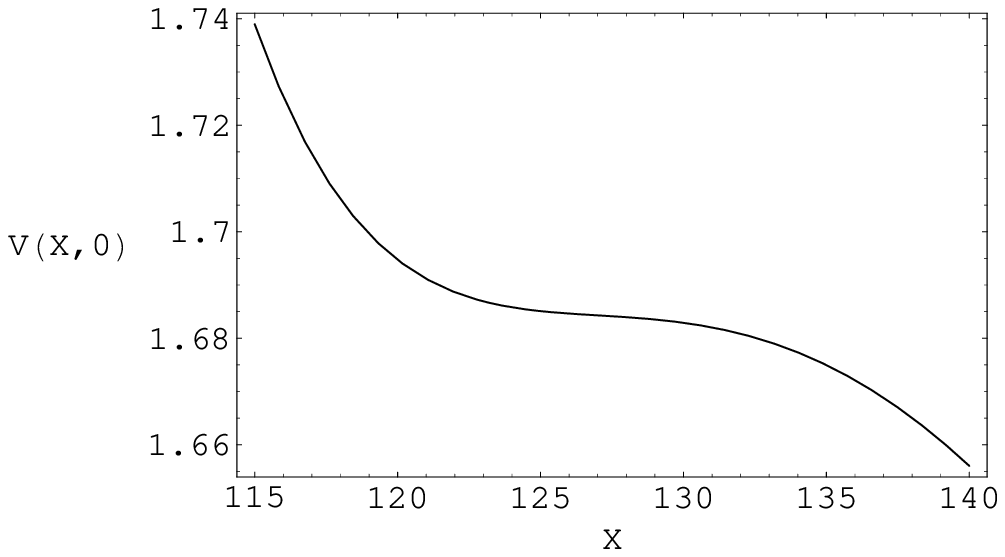}{ $Y=0$ slice of the potential for $L=60$}
\end{center}

\noindent Typically, though, we  would fix the dilaton at a value
$g_s \ll 1$, in which case $L \gg 60$. This would make it even
harder to re-establish stabilisation. Notice that this
destabilisation occurs at large radius, one where we would naively
expect to trust a classical supergravity analysis. Notice too that
the effect is sensitive to the sign of $\chi$. For the mirror
quintic, $\chi = 200 \rightarrow L = -60$, we find that
stabilisation in the $X$ direction is enhanced. However the
conditions for inflation are still destroyed since at the saddle
point we find $\eta = -2.46$.

\begin{center}
\EPSFIGURE[h]{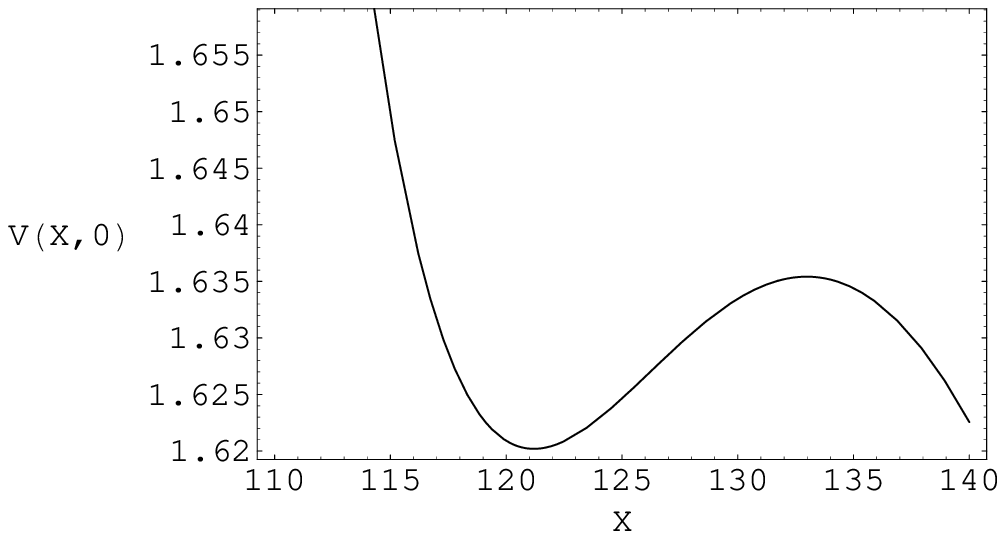}{ $Y=0$ slice of the potential for $L= - 60$. Note the enhancement of the minimum.}
\end{center}

\section{How Large is Large Enough?}

In the previous section, we saw that $\alpha'$  corrections to the
K\"ahler potential play a significant role, even at values of the
volume modulus around $100 M_P$, where one would generally expect
such corrections to be irrelevant. At what radius, then, do these
effects become negligible for determining the onset of inflation?
The answer, of course, depends on the specific compactification
manifold--in particular its Euler number, $\chi$, and on the value
of $g_s$. Below we give some examples.
To aid our analysis, we take advantage of a scaling symmetry which
the racetrack potential obeys in the absence of the correction but
which the correction term spoils. Without the correction, this
symmetry ensures that the potential and its essential properties,
such as slope and slow-roll parameters, remain the same regardless
of the location in $X$ of the saddle point. Explicitly, the
rescaling is \cite{RI}

\begin{equation}
a \rightarrow \frac{a}{\lambda}, \,\,\,\, b \rightarrow \frac{b}{\lambda}, \,\,\,\, E \rightarrow \lambda^2E
\end{equation}

\noindent
with
\begin{equation}
A \rightarrow \lambda^{3/2} A, \,\,\,\, B \rightarrow \lambda^{3/2} B, \,\,\,\, W_0 \rightarrow \lambda^{3/2} W_0
\end{equation}

\begin{equation}
X \rightarrow \lambda X, \,\,\,\, Y \rightarrow \lambda Y
\end{equation}

\noindent

With the inclusion of the $\alpha'$ corrections, this symmetry is
broken but still provides a convenient means for finding where the
large radius conclusions are spoiled.\\

\noindent In the following table we have evaluated $\eta$  for a
range of values of $\lambda$ and $L$ and compared the classical and
$\alpha'$ corrected results\footnote{ It is worth noting that not all these examples may be realised with a single K\"ahler modulus. For example, $\chi > 0 $ requires $ h^{2,1} < h^{1,1}  $. I.e. more K\"ahler moduli than complex structure moduli, which would require a rigid Calabi-Yau. In principle one could generalise this study to the case of more K\"ahler moduli, however we expect a similar effect would arise.}. The values of $L$ we have chosen
correspond to the following cases. $L=2$ corresponds to $\chi = -6$
and $g_s \approx 0.9$; this represents a value of $L$ on the low end
of physical interest. $L=60$ and $L=-60$ correspond to the quintic
and the mirror quintic respectively, with $g_s \approx 1$. $L= 2500$
corresponds to the quintic and $g_s \approx 1/12$.

\begin{center}
{\scriptsize
\begin{tabular}[h]{|c||c|c||c|c||c|c||c|c||c|c|}
\hline

 & $  L=0 $& & $  L=2 $& &$ L= 60 $& &$L=-60$& &$L=2500$&\\
 \hline
  $\lambda$ & $X_{\rm saddle}$  & $\eta$ & $X_{\rm saddle}$  & $\eta$ & $X_{\rm saddle}$  & $\eta$ &$X_{\rm saddle}$  & $\eta$&$X_{\rm saddle}$  & $\eta$  \\
 \hline

 $100 000$ &$12321632$& $-0.0061$& $12321632$& $-0.0061$& $12321632$& $-0.0061$& $12321632$&$-0.0061$&  $12321632$&$ -0.0061$\\
 $4000$ &$492865.3$  & $ -0.0061$ &$492865.3$  & $ -0.0061$ &$492865.3 $& $-0.0061 $  &$492865.2$ & $-0.0061 $&$492867.0$&$-0.0056$ \\
$400$ &$49286.53$  & $ -0.0061$ &$49286.53$  & $ -0.0061$ &$49286.66 $& $-0.0057 $  &$49286.40$ & $-0.0064 $&$49291.92$&$0.0085$ \\
$80$  &$9857.305$  & $ -0.0061$ &$9857.315$  & $-0.0059 $ &$ 9857.594$& $-0.0022 $  &$9857.017$ & $ -0.0099$&$9869.600$&$0.1582 $\\
$50$  &$6160.816$  & $ -0.0061$ &$6160.828$& $-0.0058 $  &$ 6161.182$& $0.0018 $  &$6160.451$ & $ -0.0140 $&$6176.740$&$0.3310$\\
$20$ &$2464.326$  & $ -0.0061$&$2464.346$& $-0.0050 $  &$ 2464.906$& $0.0252 $  &$2463.751$& $-0.0372 $ &$2494.857$&$1.4940$\\
$10$ &$1232.163$  & $ -0.0061$&$1232.190$& $-0.0032$  &$  1232.989$& $ 0.0827  $  &$1231.355 $ & $ -0.0935 $&no min&$-$ \\
$2$&$246.4326$  & $ -0.0061$ &$246.4938$& $0.0268$  &$ 248.5787 $& $  1.0784$  &$ 244.7928$ & $  -0.9355$&no min&$-$ \\
$1$ &$123.2163$  & $ -0.0061$ & $123.3035$ & $0.0875 $ & no min &$ - $ &$ 121.1970$ & $ -2.4635$   &no min&$-$ \\
$1/2$&$61.60816$  & $ -0.0061$&$61.73434 $& $0.2622$  & no min& $ -  $ &$59.34719$& $  -6.3833 $  &no min&$-$ \\
$1/4 $ &$30.80408$  & $ -0.0061$&  $30.99693$ &$ 0.7882$   &no min&$ - $ &$ 28.45786 $& $  -18.4781 $ &no min&$-$\\
$1/8 $ &$15.40204$  & $ -0.0061$&   no min & $ - $ &no min  & $ -$ &$12.81586 $& $ 37.7937$  &no min&$-$ \\
\hline

\end{tabular}
}
\end{center}
{{\bf Table 1:} Effect of $\alpha '$ corrections on stabilisation and  $\eta$ when we move the minimum using $\lambda$.} \label{etas }

\paragraph{}
\noindent 
Notice that for reasonable values of
$L$, say $\Re (T) \sim 10^7$ in Planck units with $2 \pi \alpha' =1$, the $\alpha'$ corrections become negligible and supergravity analysis is both qualitatively and quantitatively unaffected.
However, at smaller values of $\Re(T)$--but large enough for perturbation
theory to be justified--the
corrections not only change the details of the inflationary model but ultimately
prevent inflation from initiating. For these models, other choices of
parameters might well lead to inflation, but a lowest order calculation
is no longer adequate, even though the dimensionless expansion
parameter can be on the order of
$ \alpha' / \sqrt{T} \sim 10^{-2}$.  

\section{Conclusions}

In this note we have studied the effect that stringy corrections to
the K\"ahler potential have on K\"ahler modulus stabilisation and on
racetrack inflation. We find that $\alpha'$ corrections can play a
significant role even at values of the volume modulus  within the
perturbative realm. Explicit calculations show that the minimum
radius beyond which such corrections are irrelevant depends
sensitively on the compactification manifold and on the value of
$g_s$ (set by the fluxes). In a given model, successful
stabilization and inflation may require a sufficiently large
compactification manifold. Turning this around, we expect fluxes to
fix $W_0$ and $g_s$, the value of the cosmological constant should
fix $E$, and fundamental physics should fix the parameters of the
non-perturbative superpotential. Requiring successful moduli
stabilization and inflation may then place restrictions
on the Euler number of the compactification. 

\acknowledgments
We would like to thank Alex Hamilton, Simon Judes, Dan Kabat, Jeff Murugan and Koenraad Schalm for useful discussions throughout the course of the work. BRG acknowledges financial support from DOE grant DE-FG-02-92ER40699. The work of AW is supported in part by a NASA Graduate Student Research Fellowship grant NNG05G024H and by the Pfister Foundation.

\end{document}